# DataOps-driven CI/CD for analytics repositories

**Dmytro Valiaiev**[*]

[1]University of Arkansas Little Rock, Little Rock, Arkansas 72204, USA
**\* Corresponding author:** Dmytro Valiaiev, dvaliaiev@ualr.edu

**Abstract:** The proliferation of SQL for data processing has often occurred without the rigor of traditional software development, leading to siloed efforts, logic replication, and increased risk. This ad-hoc approach hampers data governance and makes validation nearly impossible. Organizations are adopting DataOps, a methodology combining Agile, Lean, and DevOps principles to address these challenges to treat analytics pipelines as production systems. However, a standardized framework for implementing DataOps is lacking. This perspective proposes a qualitative design for a DataOps-aligned validation framework. It introduces a DataOps Controls Scorecard, derived from a multivocal literature review, which distills key concepts into twelve testable controls. These controls are then mapped to a modular, extensible CI/CD pipeline framework designed to govern a single source of truth (SOT) SQL repository. The framework consists of five stages: Lint, Optimize, Parse, Validate, and Observe, each containing specific, automated checks. A Requirements Traceability Matrix (RTM) demonstrates how each high-level control is enforced by concrete pipeline checks, ensuring qualitative completeness. This approach provides a structured mechanism for enhancing data quality, governance, and collaboration, allowing teams to scale analytics development with transparency and control.

## 1. Introduction

Nowadays, analysts are using SQL, the de facto lingua franca of data processing, not only to pull data but also to construct entire ELT/ETL pipelines [1]. Despite its ubiquity, it is often treated as a disposable artifact lacking the rigor of traditional software development [2]. Teams work in silos, often unknowingly replicating logic that already exists elsewhere, wasting effort and increasing risk [3]. This highlights the need to treat SQL as production code and establish robust governance standards. Without a single source of truth (SOT), errors propagate exponentially. This makes reviewing the code and its validation practically impossible, even when leveraging a common repository and traceable transformations framework like dbt [4]. In such conditions, data governance becomes an afterthought.

To address these issues, businesses transition from ad hoc data analytics processes to the DataOps methodology [5]. This methodology combines Agile principles, Lean manufacturing, and DevOps [6]. DataOps practices call for treating the analytics pipeline like a production system to ensure continuous value delivery.

There is a lack of a widely standardized framework that can address efficiency, observability, and data governance in data pipelines, resulting in labor-intensive supervision and inconsistent analytical results. This perspective proposes a qualitative design for a set of DataOps‑aligned validation checks. They are meant to be embedded into a CI/CD pipeline to support data governance of an SOT SQL repository, allowing teams to collaborate, standardize, and scale analytics development with both transparency and control.

## 2. DataOps Controls Scorecard

The author conducted an extensive multivocal literature review to design a DataOps Controls Scorecard that distills key DataOps concepts into testable safeguards (Table 1). The data pipeline must satisfy these checkpoints to qualify as satisfactory. Each control captures a practice repeatedly cited across different sources. Together, they define a qualitative baseline for governed analytics development.

**Table 1.** DataOps Controls Scorecard

| # | Name | Control | Purpose |
|---|---|---|---|
| C1 | Versioning [7, 8] | All queries should be routed through a Git-based repository; any ad-hoc SQL in the data warehouse is disallowed. | Prevents drifts in logic. |
| C2 | Consistency [9, 10] | Code should have the same standard across the organization, and models should follow pre-agreed-upon naming conventions. | Maintains a uniform codebase and simplifies peer review. |
| C3 | Documentation [11, 12] | All assets should be thoroughly documented. | Promotes easier knowledge sharing and faster onboarding. |
| C4 | Ownership [13, 14] | Each model should be assigned a designated maintainer. | Promotes accountability and knowledge sharing, and prevents orphaned assets. |
| C5 | Testing [15-17] | There should be visibility and smoke-test-level controls against the existing models. | Detects defects early and safeguards data accuracy. |
| C6 | Validation [18] | Tables that are produced in the data warehouse can only result from an automated pipeline. | Prevents users from introducing scratch and temporary tables with unverified logic into production. |

**Table 1.** (*Continued*)

| # | Name | Control | Purpose |
| --- | --- | --- | --- |
| C7 | Uniqueness [19-23] | Incoming queries should serve a distinct purpose and be unique across the repository. | Prevents unnecessary data bloat and redundancy in the warehouse. |
| C8 | Performance [24-27] | Scan factors, runtime, and resource consumption should remain within allowed guardrails. | Avoids costly full-table scans and runaway jobs. |
| C9 | Automation [28-32] | Routine quality checks and governance tasks should run automatically in CI/CD workflows. | Reduces manual effort and ensures repeatable governance. |
| C10 | Observability [33-38] | Lineage tracing and a connected catalog should democratize data across the organization, with metrics surfaced in near real-time. | Provides real-time operational insight and supports compliance audits. |
| C11 | Delivery [39, 40] | All changes that pass CI/CD should be released to production as soon as possible, following the team's release cadence. | Increases speed to insight. |
| C12 | Rollback [41, 42] | In case of crisis management, the pipeline must provide the ability to undo previously submitted changes. | Enables rapid recovery during incident response. |

This ordering mirrors the life cycle of a change: authoring and standardization (C1-C4), verification (C5-C7), runtime discipline (C8-C10), rapid deployment (C11), and safe recovery (C12).

## 3. Iterative Design Approach

This review was organized using a Design-Based Research (DBR) methodology, focusing on iterative design, development, and testing of the DataOps framework solution (Figure 1). DBR combines iterative engineering with theory building. Foundational DBR work argues that useful knowledge emerges when an artifact is refined through real-world trials [43, 44]. More recently, it has been emphasized that DBR's cycles naturally accommodate stakeholder-driven adjustments [45].

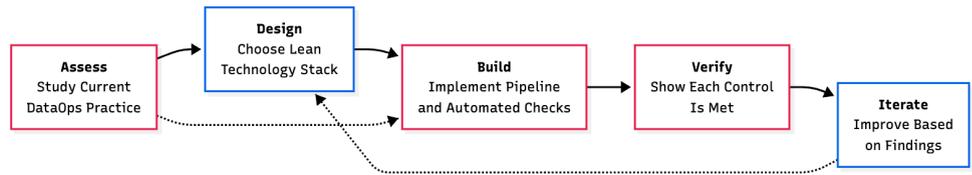

**Figure 1.** Development Lifecycle.

## 4. DataOps Validation Framework

DBR methodology is grounded in practical requirements and, therefore, necessitates a clear set of design criteria. Building on the multivocal literature review and aligned with the DataOps Controls Scorecard, the next step would be to distill key validation and deployment checks into modular components that can be integrated into an automated CI/CD pipeline. While the checks listed here represent a minimum viable set for maintaining data quality and governance, they are not exhaustive. Instead, they serve as a foundational framework that can be extended and adapted based on organizational needs. The proposed DataOps Validation Framework aligns with higher-level DataOps principles that were deduced from the literature review (Figure 2).

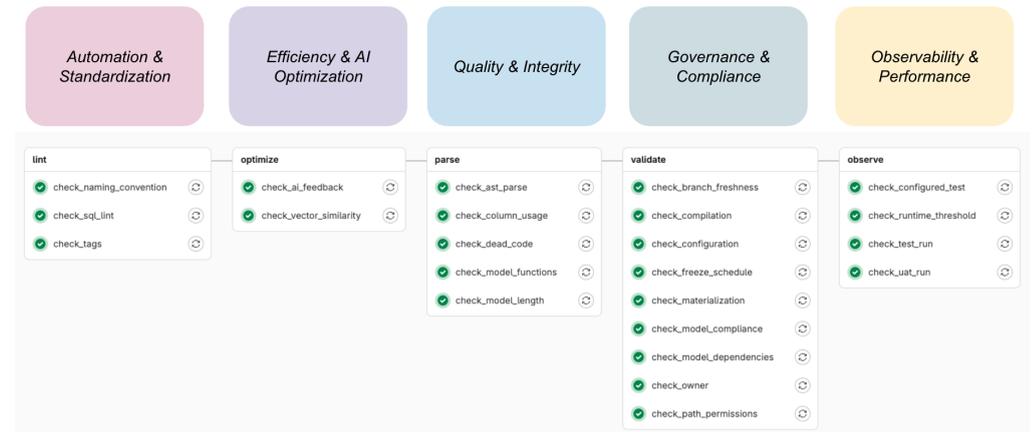

**Figure 2.** CI/CD Pipeline Graph and Respective DataOps Principles.

The validation framework described here emphasizes configurability, allowing teams to prioritize the most relevant validations and scale them across environments. The examples in Tables 2-6 are intended to be modular and reconfigurable and aligned with the principles of proactive validation, automation, and continuous improvement.

**Table 2.** Lint Stage Jobs

| # | Check | Description |
| --- | --- | --- |
| J1.1 | check_naming_convention | Ensure that all column names use snake_case. This helps distinguish raw source columns from transformed columns. The check also validates file naming conventions, simplifying ingestion into BI tools like Tableau, which automatically convert snake_case to readable headers. |

**Table 2.** (*Continued*)

| # | Check | Description |
|---|---|---|
| J1.2 | check_sql_lint | Standardize queries to follow consistent rules and formatting. Use a linter to check for syntax consistency, indentation, keyword casing, comma placement, and adherence to organizational standards. Note: code submitted to the repository becomes author-agnostic and is automatically reformatted to support comparison and uniformity across the system. |
| J1.3 | check_tags | Validate that appropriate tags are applied to the created model. Tags help assign models to the correct workflow or environment. For example, models intended for Snowflake should not be run in Redshift. Department-specific modeling (e.g., Marketing, Finance) should also follow tagging standards to maintain clarity in lineage and execution. |

**Table 3.** Optimize Stage Jobs

| # | Check | Description |
|---|---|---|
| J2.1 | check_ai_feedback | Leverage standardized prompts for an LLM to evaluate the code and catch high-level inconsistencies. This is the starting point that could be amplified by using parsed data products from the pipeline, making it aware of the underlying logic and transformations. |
| J2.2 | check_vector_similarity | Generate a TF-IDF vector and compare it to a corpus of existing models. Flag models with high cosine similarity to prevent the creation of redundant or overly similar queries. Prompt users to consolidate logic or extend existing models instead. |

**Table 4.** Parse Stage Jobs

| # | Check | Description |
|---|---|---|
| J3.1 | check_ast_parse | Extract logical components into an Abstract Syntax Tree (AST) for structural analysis and logic validation. At this stage, the query is not executed, but the analysis can catch inconsistencies, such as preventing the submission of multiple queries in a single model. The simplest check could validate the number of statements: the only option is to submit one query per file. To extend this functionality, one might parse through the entire lineage to identify the number of CTEs, subqueries, or unions. |

**Table 4.** (*Continued*)

| # | Check | Description |
|---|---|---|
| J3.2 | check_column_usage | Verify that columns referenced in intermediate steps are actually used in the final version of the model. Avoid unnecessary additions unless required for indexing or partitioning. |
| J3.3 | check_dead_code | This ensures that all pieces of code are actually being used. For instance, there can be functions or whole CTEs that do not produce any outcomes. This check would catch that. |
| J3.4 | check_model_functions | dbt relies on a staging-intermediate-marts structure. Staging models are slight transformations or subsets of incoming source data. If sources are transformed into multiple models in a different manner, disparities in the data can occur. |
| J3.5 | check_model_length | This check limits any model to a set line count, which forces a concise, single-purpose SQL. Curbing bloat speeds code reviews and keeps lineage graphs clear, improving both maintainability and warehouse performance. |

**Table 5.** Validate Stage Jobs

| # | Check | Description |
|---|---|---|
| J4.1 | check_branch_freshness | Ensure data checks are performed on the most recent data. Prevent stale branches from running by automatically failing them. Otherwise, the conflicts arise and stale changes could overwrite previously pushed fixes. |
| J4.2 | check_compilation | Ensure that the entire dbt project compiles successfully after the proposed changes. |
| J4.3 | check_configuration | Confirm model configuration is complete and valid. This way, keys that are mistyped or unsupported are not introduced into the codebase. |
| J4.4 | check_documentation | Ensure comprehensive documentation of the model, including its purpose and the data it handles. At a minimum, this includes a high-level description; ideally, documentation is also provided for each unique column to ensure downstream users understand the dataset. |

**Table 5.** (*Continued*)

| # | Check | Description |
| --- | --- | --- |
| J4.5 | check_freeze_schedule | Enforce a schedule during which code is not eligible for submission. For instance, pipelines may fail on Fridays or holidays when there is insufficient engineering support available to mitigate breakages and incidents. |
| J4.6 | check_materialization | Validate materialization strategies. For example, ephemeral should only be used for staging models, and view should not be used in the marts layer. |
| J4.7 | check_model_compliance | Detect the presence of sensitive data, including Personally Identifiable Information (PII) and Material Non-Public Information (MNPI), and ensure compliance with internal policies and external regulations (e.g., GDPR, HIPAA). Review HR and public data fields to prevent leakage or unauthorized exposure. |
| J4.8 | check_model_dependencies | Identify and flag unreferenced models, broken referential integrity, and circular or cross-dependencies. Enforce modeling best practices: flow should move from staging to intermediate to marts; marts models should not feed into intermediate layers. |
| J4.9 | check_owner | Confirm that every model has a designated owner or team. If the listed owner has left the organization, a new owner must be assigned before changes proceed. |
| J4.10 | check_path_permissions | Prevent users from submitting queries to unauthorized schemas. For example, marketing analysts should only publish models to the marketing domain. |

**Table 6.** (*Continued*)

| # | Check | Description |
| --- | --- | --- |
| J5.1 | check_configured_test | Confirms that all configured tests for models are executed correctly, verifying that models adhere to quality, performance, and correctness standards. This ensures comprehensive test coverage prior to production release. |
| J5.2 | check_runtime_threshold | Ensure that the query executes within acceptable runtime limits to avoid excessive compute usage. |

**Table 6.** Observe Stage Jobs

| # | Check | Description |
|---|---|---|
| J5.3 | check_test_run | Executes and validates configured data tests within the test environment, verifying functional correctness and alignment with predefined acceptance criteria, thus ensuring model reliability before deployment. |
| J5.4 | check_uat_run | Fully execute the given model and produce results in the testing (UAT) schema for analyst review. |

Naturally, stages that most frequently fail would require additional attention. Likewise, checks that always pass should be evaluated for potential oversimplification. This framework is designed to evolve and extend. For example, the jobs themselves can adopt a deeper taxonomy as the organization matures in data ownership.

After the model passes all respective checks, the Continuous Delivery portion of the CI/CD pipeline starts, which delivers the model (Table 7).

**Table 7.** Deployment Stage Jobs

| # | Check | Description |
|---|---|---|
| DJ1.1 | run_production | This executes the modified model in production, so all the queries and dashboards will reflect updated values as soon as possible. |
| DJ1.2 | run_documentation | This compiles all documentation and refreshes the HTML dictionary. |

This validation framework integrates the goal of implementing governance and design criteria formalized by the scorecard. It offers a structured, extensible mechanism for ensuring data product readiness through automation, modularity, and collaborative ownership.

## 5. Requirements Traceability Matrix

Each line item from the DataOps Controls Scorecard was translated into one or more concrete software-enforced checks within the DataOps Validation Framework. Next, these checks were grouped into distinct stages by a governing DataOps principle. This approach bridges high-level themes and expectations with actionable, fine-grained CI/CD enforcement. It allows the requirements to be modular and operationalized as guardrails for every new submission.

The requirements traceability matrix (RTM) maps each control to its corresponding validation mechanism and reports its enforcement status (Table 8). This RTM format provides a bidirectional link between what the system must achieve and how those requirements are met within the implemented DataOps pipeline [46, 47].

**Table 8.** Requirements Traceability Matrix

| # | Name | Control | Purpose |
| --- | --- | --- | --- |
| C1 | Versioning | check_branch_freshness, check_freeze_schedule | Verified |
| C2 | Consistency | check_sql_lint, check_naming_convention | Verified |
| C3 | Documentation | check_documentation, check_tags | Verified |
| C4 | Ownership | check_owner, check_path_permissions | Verified |
| C5 | Testing | check_configured_test, check_test_run | Verified |
| C6 | Validation | check_ast_parse, check_compilation, check_column_usage, check_dead_code, check_model_length, check_model_functions, check_model_compliance, check_materialization, check_model_dependencies, check_configuration | Verified |
| C7 | Uniqueness | check_vector_similarity | Verified |
| C8 | Performance | check_runtime_threshold | Verified |
| C9 | Automation | CI auto-trigger on commit, check_ai_feedback | Verified |
| C10 | Observability | run_documentation, HTML for dbtDocs | Supported |
| C11 | Delivery | run_production | Verified |
| C12 | Rollback | Git revert workflow | Supported |

The majority of controls are enforced by at least one CI/CD check; therefore, their status is marked as Verified. C10 (Observability) is satisfied by dbt's observability features, whereas C12 (Rollback) relies on native Git functionality [48]. Consequently, both are classified as supported.

## 6. Conclusions

Any organization's initial challenge is the absence of a centralized data infrastructure management. Before this implementation, data workflows were undocumented, siloed, and governed by tribal knowledge. The successful deployment of an automated system not only replaced these ad hoc practices but also introduced rigorous controls around documentation, ownership, testing, and delivery.

The findings of this review highlight the feasibility of treating data engineering infrastructure as code. The proposed framework is designed to promote reusability and modularity in data pipelines. Each check has been implemented as a standalone module, offering flexibility for reuse in different pipelines. Moreover, these modules can be easily toggled on or off when testing, evolving, or redesigning pipeline components. Such modular pipelines represent a straightforward architecture that can evolve in parallel as the organization reaches a higher level of data maturity. A requirements traceability matrix (RTM) mapped every item on the DataOps Controls Scorecard to its corresponding pipeline implementation component, demonstrating the design's qualitative completeness.